\documentclass[preprint,3p,times]{elsarticle}

\usepackage{amssymb}

\usepackage{amsmath}

\usepackage{hyperref} 
\usepackage{marginnote}
\usepackage{xcolor}
\usepackage{amsthm}
\usepackage{float}
\usepackage{tikz}
\usepackage{booktabs}
\usepackage{pgfplots}
\pgfplotsset{compat = 1.3}
\usepackage{mathtools}
\usepackage{caption}
\usepackage{subcaption}

\biboptions{numbers,sort&compress}

\numberwithin{equation}{section}
\numberwithin{figure}{section}
\numberwithin{table}{section}

\newcommand{\C}{\mathbb{C}}
\newcommand{\R}{\mathbb{R}}
\newcommand{\cay}{\mathrm{cay}}
\newcommand{\caymod}{\widetilde{\mathrm{cay}}}
\theoremstyle{plain}
\newtheorem{theorem}{Theorem}[section]
\newtheorem{lemma}[theorem]{Lemma}
\theoremstyle{definition}
\newtheorem{remark}[theorem]{Remark}
\newtheorem{example}[theorem]{Example}
\newtheorem{definition}[theorem]{Definition}

\definecolor{fk4blue}{rgb}{0.00000,0.46275,0.68627}

\journal{}

\begin{document}

\begin{frontmatter}



\title{A modified Cayley transform for SU(3) molecular dynamics simulations}

\author[1]{Kevin Schäfers\texorpdfstring{\corref{cor1}}{}}
\ead{schaefers@math.uni-wuppertal.de}
\author[2]{Michael Peardon}
\ead{mjp@maths.tcd.ie}
\author[1]{Michael Günther}
\ead{guenther@math.uni-wuppertal.de}

\cortext[cor1]{Corresponding author.}

\affiliation[1]{organization={Institute of Mathematical Modelling, Analysis and Computational Mathematics (IMACM), Chair of Applied and Computational Mathematics, Bergische Universität Wuppertal},
            addressline={Gaußstraße 20}, 
            city={Wuppertal},
            postcode={42119}, 
            country={Germany}}
\affiliation[2]{organization={School of Mathematics, Trinity College Dublin},
            city={Dublin 2},
            country={Ireland}}

\begin{abstract}
We propose a modification to the Cayley transform that defines a suitable local parameterization for the special unitary group SU(3). 
The new mapping is used to construct splitting methods for separable Hamiltonian systems whose phase space is the cotangent bundle of SU(3) or, more general, $\text{SU(3)}^N,\ N \in \mathbb{N}$. Special attention is given to the hybrid Monte Carlo algorithm for gauge field generation in lattice quantum chromodynamics. We show that the use of the modified Cayley transform instead of the matrix exponential neither affects the time-reversibility nor the volume-preservation of the splitting method. Furthermore, the advantages and disadvantages of the Cayley-based algorithms are discussed and illustrated in pure gauge field simulations.
\end{abstract}

\end{frontmatter}


\section{Introduction}
The Hybrid Monte Carlo (HMC) algorithm \cite{duane1987hybrid} is a frequent choice for gauge field generation in lattice quantum chromodynamics (QCD). 
In the molecular dynamics step of the HMC algorithm, Hamiltonian equations of motion have to be solved using a volume-preserving and time-reversible numerical integration scheme in order to satisfy the detailed balance condition, ensuring that the equilibrium distribution of the Markov chain can be reached \cite{knechtli2017lattice}.
Moreover, the differential equation for the link variables is of Lie-type, i.e.\ the exact solution to this equation evolves on a Lie group manifold, in particular the links are elements of the special unitary group $\mathrm{SU}(3)$.
In order to avoid non-physical numerical approximations, we demand that the numerical scheme satisfies the closure property, i.e.\ it yields numerical approximations situated on the Lie group manifold.
Due to the separability of the Hamiltonian, splitting methods \cite{mclachlan2002splitting} provide a tool to solve the equations of motion explicitly while preserving the time-reversibility, symplecticity, as well as the closure property of the exact flow. By computing the exact flows of the respective subsystems, the matrix exponential $\exp(A) \coloneqq \sum\nolimits_{k \ge 0} A^k/k!$ maps elements of the Lie algebra $\mathfrak{su}(3)$ into the Lie group $\mathrm{SU}(3)$. In the context of Lie group integrators, the exponential map is a natural choice as a local parameterization $\Psi \colon \mathfrak{g} \to \mathcal{G}$. An alternative local parameterization for quadratic Lie groups of the form
\begin{align}\label{eq:quadratic_Lie-group}
\mathcal{G} = \{ Y \in \mathrm{GL}(n) \,\vert\, Y^\dag J Y = J\}, \quad J \in \mathbb{R}^{n \times n} \; \mathrm{const.},
\end{align} 
is given by the Cayley transform \cite{Diele1998Cayley,Diele1998Cayley2,LOPEZ2001Cayley,Iserles2001Cayley,MARTHINSEN2001Cayley,HairerLubichWanner}
\begin{align}\label{eq:cayley-transform}
    \cay(\Omega) =(I - \Omega)^{-1} (I + \Omega).
\end{align}
The Cayley transform is also feasible for the special unitary group $\mathrm{SU}(2)$. Recently, the Cayley transform has been applied successfully for simulation in an $\mathrm{SU}(2)$ Yang-Mills theory using the HMC algorithm \cite{wandelt2021geometric}, resulting in a significantly more efficient computational process. Unfortunately, the Cayley transform does not define a local parameterization for $\mathrm{SU}(3)$.
This paper contributes in providing a modification to the Cayley transform that defines a local parameterization for the special unitary group $\mathrm{SU}(3)$ and investigates its use in lattice QCD. 

In Section \ref{sec:cayley_overview}, we will briefly explain why the Cayley transform is a suitable local parameterization for $\mathrm{SU}(2)$, but not for $\mathrm{SU}(3)$ anymore.
Then, in Section \ref{sec:modified_cayley}, we introduce a modified Cayley transform that defines a local parameterization from $\mathfrak{su}(3)$ into $\mathrm{SU}(3)$. 
In Section \ref{sec:splitting_methods}, we consider splitting methods for lattice QCD simulations and then show how to use the modified Cayley transform inside these methods. Here, we show that the usage of the modified Cayley transform does not affect the time-reversibility and volume-preservation. Moreover, we will investigate the convergence order of the resulting integrators. 
The proposed integrators are tested by simulating lattice gauge fields and compared to integrators based on the exponential map in Section \ref{sec:numerical_results}. 
The paper closes by some concluding remarks and outlook for future research.

\section{The Cayley transform and the special unitary group}\label{sec:cayley_overview}
The Cayley transform \eqref{eq:cayley-transform} defines a local parameterization for quadratic Lie groups of the form \eqref{eq:quadratic_Lie-group}, i.e.\ it maps elements from the Lie algebra $$\mathfrak{g} = \{ \Omega \in \mathbb{R}^{n \times n} \, \vert \, J\Omega + \Omega^\dag J = 0\}$$ into $\mathcal{G}$. Furthermore, it is a local diffeomorphism near $\Omega = 0$ \cite[Lemma IV.8.7]{HairerLubichWanner}. Prominent examples of quadratic Lie groups are the orthogonal group $\mathrm{O}(n)$ and the unitary group $\mathrm{U}(n)$. The special orthogonal group $\mathrm{SO}(n)$ and the special unitary group $\mathrm{SU}(n)$ are not of the form \eqref{eq:quadratic_Lie-group}. 
The Cayley transform also defines a mapping from $\mathfrak{so}(n)$ into $\mathrm{SO}(n)$. Since the eigenvalues of any skew-symmetric matrix $\Omega \in \mathfrak{so}(n)$ are purely imaginary and complex eigenvalues of real-valued matrices occur in complex conjugate pairs, it can be shown that 
$$ \det \left( (I-\Omega)^{-1} (I+\Omega) \right) = 1 \quad \forall \Omega \in \mathfrak{so}(n),$$
i.e., $\cay(\Omega) \in \mathrm{SO}(n)$ for all $\Omega \in \mathfrak{so}(n)$. 
In general, eigenvalues of complex-valued matrices do not necessarily occur in complex conjugate pairs. 
For elements of the Lie algebra $\mathfrak{su}(2)$, however, this is still true, i.e., the Cayley transform \eqref{eq:cayley-transform} also defines a mapping from $\mathfrak{su}(2)$ to $\mathrm{SU}(2)$. 
Unfortunately, the situation changes for elements of $\mathfrak{su}(3)$.

\lemma[Cayley transform for $\mathrm{SU}(3)$]{The Cayley transform \eqref{eq:cayley-transform} maps elements from the Lie algebra $\mathfrak{su}(3)$ into the Lie group $\mathrm{U}(3)$, but not necessarily into $\mathrm{SU}(3)$.}
\proof{Since the Lie algebra $\mathfrak{su}(3)$ is a subalgebra of $\mathfrak{u}(3)$ and $\mathrm{U}(3)$ is of the form \eqref{eq:quadratic_Lie-group}, the Cayley transform definitely maps elements from $\mathfrak{su}(3)$ into $\mathrm{U}(3)$. It remains to show that the Cayley transform does not necessarily map into the subgroup $\mathrm{SU}(3)$.
Any matrix $\Omega \in \mathfrak{su}(3)$ can be represented by $\Omega = \sum\nolimits_{j=1}^8 c_j i \lambda_j$ with real-valued coefficients $c_j$ and $\lambda_j$ the Gell-Mann matrices (see \ref{app:Gell-Mann}). Applying the Cayley transform to the basis element $i\lambda_8$ yields 
\begin{align*}
    \cay(i\lambda_8) = \mathrm{diag}\begin{pmatrix}
        \tfrac{1}{2} + i \tfrac{\sqrt{3}}{2} & \tfrac{1}{2} + i \tfrac{\sqrt{3}}{2} & - \tfrac{1}{7} - i \tfrac{4\sqrt{3}}{7}
    \end{pmatrix}
\end{align*}
with $\det(\cay(i\lambda_8)) = \frac{13}{14} + i \frac{3\sqrt{3}}{14} \neq 1$, concluding the proof. \qedhere}

\section{Modified Cayley transform for SU(3)}\label{sec:modified_cayley}
Consider a matrix $M \in \C^{3 \times 3}$. Then, the Cayley--Hamilton theorem implies
\begin{align}\label{eq:CayleyHamilton}
    M^3 - \mathrm{tr}(M) M^2 + \tfrac{1}{2} \left( \mathrm{tr}(M)^2 - \mathrm{tr}(M^2) \right) M - \det(M) I = 0.
\end{align}
Taking the trace of this matrix expression yields 
\begin{align}\label{eq:FormulaDet}
\begin{split}
    \det (M) &= \tfrac{1}{3} \mathrm{tr}(M^3) - \tfrac{1}{3} \mathrm{tr}(M) \mathrm{tr}(M^2) + \tfrac{1}{6} \mathrm{tr}(M)^3 - \tfrac{1}{6} \mathrm{tr}(M) \mathrm{tr}(M^2) = \tfrac{1}{3} \mathrm{tr}(M^3) - \tfrac{1}{2} \mathrm{tr} (M) \mathrm{tr}(M^2) + \tfrac{1}{6} \mathrm{tr}(M)^3.
\end{split}
\end{align}
For $M=I+\Omega$ with $\Omega$ being traceless, \eqref{eq:FormulaDet} simplifies to
\begin{align}\label{FormulaDet2}
\begin{split}
    \det(I+\Omega) &= \tfrac{1}{3} \mathrm{tr}\left( (I+\Omega)^3 \right) - \tfrac{1}{2} \mathrm{tr}\left( I+\Omega\right) \mathrm{tr} \left( (I+\Omega)^2\right) + \tfrac{1}{6} \mathrm{tr} (I + \Omega)^3 
    = 1 - \tfrac{1}{2} \mathrm{tr} (\Omega^2) + \tfrac{1}{3} \mathrm{tr} (\Omega^3) 
    \\
    &= 1 - \tfrac{1}{2} \mathrm{tr} (\Omega^2) + \det(\Omega),
\end{split}
\end{align}
where we applied \eqref{eq:FormulaDet} to the traceless matrix $\Omega$ in the last step. 

\lemma{If $\Omega \in \C^{3 \times 3}$ is an anti-Hermitian matrix, then for any angle $\theta \in (- \pi, \pi]$, the matrix 
\begin{align}\label{modCayley}
    U(\Omega) = \left( I - e^{-i \theta} \Omega\right)^{-1} \left( I + e^{i \theta} \Omega\right)
\end{align}
obeys $U(\Omega)^\dag U(\Omega) = I$ and thus is an element of the Lie group $\mathrm{U}(3)$.}
\proof{
By using the commutativity of the matrices and that $\Omega^\dagger = -\Omega$ holds, we obtain
\begin{align*}
    U(\Omega)^\dag U(\Omega) &= (I + e^{i \theta}\Omega)^\dag \left((I - e^{-i\theta} \Omega)^{-1}\right)^{\dag} (I - e^{-i\theta} \Omega)^{-1} (I+e^{i\theta}\Omega) = (I - e^{-i\theta}\Omega) (I + e^{i\theta}\Omega)^{-1} (I-e^{-i\theta}\Omega)^{-1} (I + e^{i\theta} \Omega) = I.
\end{align*}
\qedhere
}

\remark{For $\theta =0$, \eqref{modCayley} reduces to the traditional Cayley transform since $e^{-i\theta} \vert_{\theta = 0} = e^{i\theta}\vert_{\theta = 0} = 1$.}\medskip\normalfont

In a next step, we will determine the angle $\theta$ so that $\det(U(\Omega)) = 1$, i.e.\ the transformation \eqref{modCayley} maps elements $\Omega \in \mathfrak{su}(3)$ into $\mathrm{SU}(3)$.

\lemma{\label{detCondition} \normalfont
Consider the matrix $U$ given by \eqref{modCayley} and $\Omega \in \mathfrak{su}(3)$. It holds
\begin{align*}
    \det(U(\Omega)) = 1 \quad \Leftrightarrow \quad \det\left(I + e^{i\theta}\Omega\right) \in \R.
\end{align*}
}
\proof{
With $\det(U(\Omega)) = \det\left(I + e^{i\theta} \Omega\right)/\det\left(I - e^{-i\theta} \Omega\right)$,
it holds 
\begin{align*}
    \det(U(\Omega)) = 1 \quad \Leftrightarrow \quad \det\left( I + e^{i\theta} \Omega \right) = \det \left( I - e^{-i\theta} \Omega\right).
\end{align*}
By applying \eqref{FormulaDet2}, we obtain
\begin{align*}
    1 + \det\left(e^{i\theta}\Omega\right) - \tfrac{1}{2} \mathrm{tr}\left(e^{2i\theta}\Omega^2\right) &=1 + \det\left(-e^{-i\theta}\Omega\right) - \tfrac{1}{2} \mathrm{tr}\left(e^{-2i\theta}\Omega^2\right) \\
    \Leftrightarrow \qquad\quad e^{3i\theta}\det\left(\Omega\right) - \frac{e^{2i\theta}}{2} \mathrm{tr}\left(\Omega^2\right) &= -e^{-3i\theta}\det\left(\Omega\right) - \frac{e^{-2i\theta}}{2} \mathrm{tr}\left(\Omega^2\right).
\end{align*}
Applying Euler's formula then yields 
\begin{align*}
    \cos(3\theta)\det\left(\Omega\right) - \tfrac{1}{2} i\sin(2\theta) \mathrm{tr}\left(\Omega^2\right)
    &=  -\cos(3\theta)\det\left(\Omega\right) + \tfrac{1}{2} i\sin(2\theta) \mathrm{tr}\left(\Omega^2\right).
\end{align*}
Since the determinant of $\Omega \in \mathfrak{su}(3)$ is purely imaginary and $\mathrm{tr}\left(\Omega^2\right) \in \R$, we can rewrite the expression and obtain 
\begin{align*}
    & & i\cos(3\theta)\Im(\det(\Omega)) - \tfrac{1}{2} i\sin(2\theta) \mathrm{tr}\left(\Omega^2\right)
    &=  -i\cos(3\theta)\Im(\det(\Omega)) + \tfrac{1}{2} i\sin(2\theta) \mathrm{tr}\left(\Omega^2\right) \\
    &\Leftrightarrow& \cos(3\theta)\Im(\det(\Omega)) - \tfrac{1}{2} \sin(2\theta) \mathrm{tr}\left(\Omega^2\right)
    &=  -\cos(3\theta)\Im(\det(\Omega)) + \tfrac{1}{2}  \sin(2\theta) \mathrm{tr}\left(\Omega^2\right).
\end{align*}
The left-hand side and the right-hand side coincide with the imaginary part of $\det\left(I + e^{i\theta} \Omega\right)$ and $\det\left(I - e^{-i\theta}\Omega\right)$, respectively. Since $\Im{\left(\det\left(I + e^{i\theta}\Omega\right)\right)} = - \Im{\left(\det\left(I - e^{-i\theta}\Omega\right)\right)}$, the only solution is given by
\begin{align*}
    \Im{\left(\det\left(I + e^{i\theta}\Omega\right)\right)} = 0 = \Im{\left(\det\left(I - e^{-i\theta}\Omega\right)\right)},
\end{align*}
which concludes the proof. \qedhere
}\medskip

\theorem{
The angle 
\begin{align*}
    \theta(\Omega) \coloneqq \sin^{-1} \left(-\frac{1}{2} \left( \gamma(\Omega)^{-1} \mp \sqrt{\gamma(\Omega)^{-2} + 1}\,\right)\right),
\end{align*}
with $\gamma(\Omega) \coloneqq 4 \Im(\det(\Omega)) / \mathrm{tr}(\Omega^2)$,
ensures $\det\left(U(\Omega)\right) = 1$.
}\proof{
Applying \eqref{FormulaDet2} and Euler's formula yields
\begin{align*}
    \det\left( I + e^{i\theta} \Omega\right) &= 1 + e^{3i\theta}\det\left(\Omega\right) - \tfrac{1}{2} e^{2i\theta} \mathrm{tr}\left(\Omega^2\right) = 1 + \left( \cos(3 \theta) + i \sin(3 \theta)\right) \det\left(\Omega\right) - \tfrac{1}{2} \left( \cos(2\theta) + i \sin(2\theta)\right) \mathrm{tr}\left(\Omega^2\right).
\end{align*}
According to Lemma \ref{detCondition}, we demand $\Im(\det( I + e^{i\theta} \Omega)) = 0$, i.e.,
\begin{align*}
    0 = \cos(3\theta) \Im(\det(\Omega)) - \tfrac{1}{2} \sin(2 \theta) \mathrm{tr}\left( \Omega^2 \right) \quad
    \Leftrightarrow \quad \tfrac{1}{2} \sin(2 \theta) \mathrm{tr}\left( \Omega^2 \right) = \cos(3\theta) \Im(\det(\Omega)).
\end{align*}
Using the double- and triple-angle formula results in an equivalent expression which is given by
\begin{align*}
    \sin(\theta)\cos(\theta) \mathrm{tr}\left( \Omega^2 \right) &= \cos(\theta) \left(1 - 4\sin^2(\theta)\right) \Im(\det(\Omega)).
\end{align*}
This gives the trivial result $\cos(\theta) = 0$, for which $U=I \in \mathrm{SU}(3)$, as well as a non-trivial constraint
\begin{align*}
    \sin(\theta) \mathrm{tr}\left(\Omega^2\right) = \left(1 - 4\sin^2(\theta)\right) \Im(\det(\Omega)).
\end{align*}
For $\Im(\det(\Omega)) = 0$, $\sin(\theta)=0$ solves this equation. For $\Im(\det(\Omega)) \neq 0$, we obtain
\begin{align*}
    \sin(\theta) \mathrm{tr}\left(\Omega^2\right) = (1 - 4\sin^2(\theta)) \Im(\det(\Omega)) 
    \quad \Leftrightarrow \quad \sin^2(\theta) + \frac{\mathrm{tr}(\Omega^2)}{4 \Im(\det(\Omega))} \sin(\theta) - \frac{1}{4} = 0.
\end{align*}
Solving this equation for $\sin(\theta)$ yields 
\begin{align*}
    \sin(\theta) &= - \frac{1}{2} \left( \frac{\mathrm{tr}(\Omega^2)}{4 \Im(\det(\Omega))} \mp \sqrt{\left( \frac{\mathrm{tr}(\Omega^2)}{4 \Im(\det(\Omega))} \right)^2 + 1}\; \right).
\end{align*}
By introducing the parameter $\gamma(\Omega) \coloneqq 4\Im(\det(\Omega))/ \mathrm{tr}(\Omega^2)$, the angle $\theta$ is given by
\begin{align*}
    \theta(\Omega) &= \sin^{-1} \left(-\frac{1}{2} \left( \gamma(\Omega)^{-1} \mp \sqrt{\gamma(\Omega)^{-2} + 1}\right)\right). \qedhere
\end{align*}}\normalfont

\begin{figure}
    \centering
    \includegraphics[]{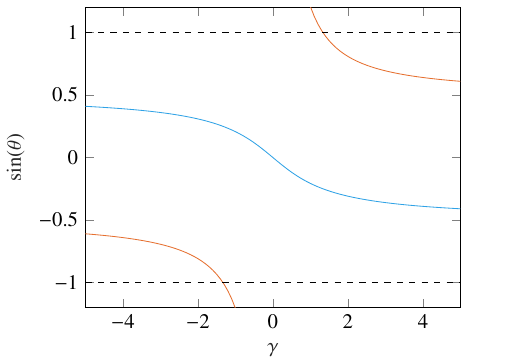}
    \caption{The modified Cayley phase.}
    \label{fig:sin}
\end{figure}

Due to the quadratic equation for $\sin(\theta)$, there are two solutions in general. They are depicted in Figure \ref{fig:sin}. One branch (shown in red) diverges at $\gamma = 0$ and for small values does not give a real phase (denoted by the black dashed lines). If we exclude this branch, the phase is single-valued, differentiable for all $\gamma$ and bounded such that $\sin(\theta) \in \left( -1/2,1/2 \right)$ (shown in blue). The functional form of the remaining branch is given by 
\begin{align*}
    \sin(\theta) = \begin{cases}
    0, & \gamma(\Omega) = 0, \\
    -\tfrac{1}{2} \left( \gamma(\Omega)^{-1} - \sqrt{\gamma(\Omega)^{-2} + 1}\right), & \gamma(\Omega) > 0, \\
    -\tfrac{1}{2} \left( \gamma(\Omega)^{-1} + \sqrt{\gamma(\Omega)^{-2} + 1}\right), & \gamma(\Omega) < 0.
    \end{cases}
\end{align*}

\noindent By applying $\sin^{-1}$ on both sides and choosing the branch $\theta \in (-\pi/6,\pi/6)$, we have $U(-\Omega) = U(\Omega)^\dag$ as the following lemma shows.

\lemma{
Taking the angle $\theta \in (-\pi/6,\pi/6)$ ensures $\theta(-\Omega) = - \theta(\Omega)$, which in turn makes sure that the crucial property $U(-\Omega) = U(\Omega)^\dag$ holds.
}
\proof{
Without loss of generality, assume $\gamma(\Omega) > 0$. Then, $\gamma(-\Omega) = -\gamma(\Omega) < 0$ and thus 
\begin{align*}
    \theta(-\Omega) &= \sin^{-1} \left( - \frac{1}{2} \left(-\gamma(\Omega)^{-1} + \sqrt{\gamma(\Omega)^{-2} + 1} \right) \right) 
    = \sin^{-1} \left( \frac{1}{2} \left(\gamma(\Omega)^{-1} - \sqrt{\gamma(\Omega)^{-2} + 1} \right) \right) \\
    &= -\sin^{-1} \left( -\frac{1}{2} \left(\gamma(\Omega)^{-1} - \sqrt{\gamma(\Omega)^{-2} + 1} \right) \right) 
    = -\theta(\Omega).
\end{align*}
Consequently, it holds 
\begin{align*}
    U(-\Omega) = (I + e^{i\theta(\Omega)} \Omega)^{-1} (I - e^{-i\theta(\Omega)} \Omega),
\end{align*}
which is identical to 
\begin{align*}
    U(\Omega)^\dag = (I-e^{-i\theta(\Omega)} \Omega) (I+ e^{i\theta(\Omega)}\Omega)^{-1},
\end{align*}
due to the commutativity of the matrices. \qedhere
}\medskip

\noindent Based on these results, we are now able to define the modified Cayley transform for SU$(3)$.

\definition[Modified Cayley transform for SU$(3)$]{
Let $\Omega \in \mathfrak{su}(3)$. With 
\begin{align*}
    \theta(\Omega) \coloneqq \begin{cases}
    0, & \gamma(\Omega) = 0, \\
    \sin^{-1} \left( -\tfrac{1}{2} \left( \gamma(\Omega)^{-1} - \sqrt{\gamma(\Omega)^{-2} + 1}\right)\right), & \gamma(\Omega) > 0, \\
    \sin^{-1} \left( -\tfrac{1}{2} \left( \gamma(\Omega)^{-1} + \sqrt{\gamma(\Omega)^{-2} + 1}\right)\right), & \gamma(\Omega) < 0,
    \end{cases}
\end{align*} 
and $\gamma(\Omega) \coloneqq 4 \Im(\det(\Omega))/\mathrm{tr}(\Omega^2)$ so that $\theta(\Omega) \in (-\pi/6,\pi/6)$, the modified Cayley transform
\begin{align}\label{modified Cayley}
    \widetilde{\cay}(\Omega) \coloneqq \left( I - e^{-i\theta(\Omega)} \Omega\right)^{-1} \left( I + e^{i \theta(\Omega)} \Omega \right),
\end{align}
maps elements from $\mathfrak{su}(3)$ into $\mathrm{SU}(3)$. Moreover, it is a local diffeomorphism near $\Omega = 0$.
}\normalfont\medskip

The modified Cayley transform \eqref{modified Cayley} provides an alternative local parameterization for numerical integration on the special unitary group $\mathrm{SU}(3)$. 

\remark[Implementation of the modified Cayley transform]{
In practical computations, one can considerably simplify the computation of $e^{i \theta} = \cos(\theta) + i \sin(\theta)$ and $-e^{-i\theta} = -\cos(\theta) + i \sin(\theta)$ via
$$ \sin(\theta) = - \frac{1}{2} \left(\gamma(\Omega)^{-1} \mp \sqrt{\gamma(\Omega)^{-2} + 1}\right), \qquad \cos(\theta) = \sqrt{1 - \sin(\theta)^2}. $$
The inverse of $X := (I - e^{-i\theta(\Omega)}\Omega)$ can be computed via
\mbox{$X^{-1} = \mathrm{adj}(X)/\det(X)$}.
}\normalfont\medskip%

Next, we will investigate the use of the modified Cayley transform inside splitting methods to solve the Hamiltonian equations of motion occurring in the molecular dynamics step of the HMC algorithm for lattice QCD computations in $\mathrm{SU}(3)$.

\section{Lattice QCD simulations using the modified Cayley transform}\label{sec:splitting_methods}
In this section, we focus on lattice QCD simulations on a $d$-dimensional lattice of size $V = T \times L^{d-1}$ where one considers the separable Hamiltonian system 
\begin{align}\label{eq:Hamiltonian}
    H([P],[U]) = \frac12 \sum\limits_{x,\mu} \mathrm{tr}(P_{x,\mu}^2) + S([U]),
\end{align}
with kinetic energy $\langle P,P\rangle / 2$ and $S([U])$ denoting the potential energy/action term. Here, the links $U_{x,\mu}$, connecting the sites $x$ and $x + a \hat{\mu}$, are elements of the special unitary group $\mathrm{SU}(3)$, and the scaled momenta $iP_{x,\mu}$ are elements of the associated Lie algebra $\mathfrak{su}(3)$. One can express the Lie algebra elements as $iP_{x,\mu} = p_{x,\mu}^{j} T_{j}$ where we choose normalized generators $T_{j} \coloneqq \sqrt{-a/2} i \lambda_{j}$ with $\lambda_{j}$ the Gell-Mann matrices \eqref{eq:Gell-Mann} and $a$ the lattice spacing. The generators are linked to the Lie group elements via the right-invariant differential operator $e_{j}$ whose action on $U_{x,\mu}$ is defined by $e_{j}(U_{x,\mu}) = - T_{j} U_{x,\mu}$. Consequently, the equations of motion read 
\begin{subequations}\label{eq:equations_of_motion}
    \begin{align}
        i\dot{P}_{x,\mu} &= \left\{- e_{j}(S([U])) T^{j}\right\}_{x,\mu}, \label{eq:momentum_ODE}\\
        \dot{U}_{x,\mu} &= -iP_{x,\mu} U_{x,\mu}. \label{eq:link_ODE}
    \end{align}
\end{subequations}
Since the Hamiltonian \eqref{eq:Hamiltonian} is separable, splitting methods \cite{mclachlan2002splitting} provide a tool to efficiently integrate the equations of motion \eqref{eq:equations_of_motion} while preserving the time-reversibility, symplecticity, as well as the closure property of the exact flow.

\subsection{Splitting methods}
Splitting methods are a framework for efficiently solving initial-value problems of ordinary differential equations (ODEs)
$$\dot{y} = f(y), \quad y(t_0) = y_0,$$
by splitting the system into (in general $N \in \mathbb{N}$) subsystems 
$$\dot{y} = f^{\{q\}}(y), \quad \sum\nolimits_{q=1}^N f^{\{q\}}(y) = f(y),$$ 
that are easier to solve than the overall system. Ideally, one is able to compute the exact flows $\varphi_t^{\{q\}}(y_0)$ to the respective subsystems $\dot{y} = f^{\{q\}}(y),\; y(t_0)=y_0$. Splitting methods then compute a numerical approximation to the overall ODE system by composing evaluations of the exact flows to the respective subsystems. 
For $N=2$, one step of a splitting method with step size $h$ reads
\begin{align*}
    \Phi_h(y_0) = \varphi_{b_s h}^{\{2\}} \circ \varphi_{a_s h}^{\{1\}} \circ \varphi_{b_{s-1} h}^{\{2\}} \circ \ldots \circ \varphi_{a_2 h}^{\{1\}} \circ \varphi_{b_1 h}^{\{2\}} \circ \varphi_{a_1 h}^{\{1\}}(y_0),
\end{align*}
where the weights $a_j,b_j \in \mathbb{R}$ have to satisfy the order-1 conditions $\sum\nolimits_j a_j = \sum\nolimits_j b_j = 1$ for convergence, and the symmetry conditions 
\begin{align*}
    a_1 &= 0, & a_j &= a_{s+2-j}, & b_j &= b_{s+1-j} & &\text{(velocity version)},
\intertext{or}
    b_s &= 0, & b_j &= b_{s-j}, & a_j &= a_{s+1-j} & &\text{(position version)},
\end{align*}
to obtain a time-reversible scheme.

In lattice QCD, where the Hamiltonian \eqref{eq:Hamiltonian} is separable, i.e.\ it takes the form $H(p,q) = T(p) + V(q)$, we consider the two-way partitioned system 
\begin{align*}
    \begin{pmatrix}
        i \dot{P}_{x,\mu} \\ \dot{U}_{x,\mu}
    \end{pmatrix} &= \begin{pmatrix}
        0 \\ - iP_{x,\mu} U_{x,\mu}
    \end{pmatrix} + \begin{pmatrix}
        \left\{- e_{j}(S([U])) T^{j}\right\}_{x,\mu} \\ 0
    \end{pmatrix}.
\end{align*}
The exact flows of the two subsystems can be computed analytically and are given by 
\begin{subequations}\label{eq:exact_flows}
\begin{align}
    \varphi_t^{\{1\}}([P],[U]) &= \begin{pmatrix}
        iP_{x,\mu} \\ \exp(-tiP_{x,\mu}) U_{x,\mu}
    \end{pmatrix}, \label{eq:link_update}
\intertext{and}
\varphi_t^{\{2\}}([P],[U]) &= \begin{pmatrix}
        P_{x,\mu} - t \left\{e_{j}(S([U]))T^{j}\right\}_{x,\mu} \\ U_{x,\mu}
    \end{pmatrix}, \label{eq:momentum_update}
\end{align}
\end{subequations}
respectively. Note that the momentum update \eqref{eq:momentum_update} can be regarded as an forward Euler step of step size $t$ applied to the first subsystem. Similarly, one can regard the link update \eqref{eq:link_update} as one time step of the Munthe-Kaas approach \cite{munthe1998runge} based on the forward Euler method, usually known as the \textit{Lie-Euler method}.

\example[Lie-Euler method for $\mathrm{SU}(3)$]{
Consider the ODE of Lie-type
\begin{align*}
    \dot{Y}(t) = A(t) \cdot Y(t), \quad Y(t_0) = Y_0 \in \mathrm{SU}(3),
\end{align*}
with $A(t) \in \mathfrak{su}(3) \; \forall t$. 
Let $A,\Omega \in \mathfrak{su}(3)$ and $\Psi \colon \mathfrak{su}(3) \to \mathrm{SU}(3)$ denote a local parameterization. 
The differential of the local parameterization $\Psi$ is given as
$$\left(\frac{\mathrm{d}}{\mathrm{d} \Omega} \Psi(\Omega) \right)A = \mathrm{d}\Psi_{\Omega}(A) \Psi(\Omega)$$
where $\mathrm{d}\Psi_{\Omega} \colon \mathfrak{su}(3) \to \mathfrak{su}(3)$.
Then, a numerical approximation $Y_1 \approx Y(t_1)$ at time point $t_1 = t_0 + h$ is obtained by
\begin{align}\label{eq:Lie-Euler}
    Y_1 = \Psi\left(h \mathrm{d} \Psi_{\Omega}^{-1}(A(t_0)) \big\vert_{\Omega = 0}\right) Y_0,
\end{align}
where $\mathrm{d}\Psi_{\Omega}^{-1}$ denotes the inverse of the operator $\mathrm{d}\Psi_{\Omega}$.
For $\Psi = \exp$, the update becomes 
\begin{align*}
    Y_1 = \exp\left(h A(t_0)\right) Y_0.
\end{align*}
}\normalfont

Using another local parameterization in \eqref{eq:Lie-Euler} provides an order-1 approximation to the exact flow \eqref{eq:link_update} and, depending on the particular parameterization, may result in a more efficient computational process. In \cite{wandelt2021geometric}, this idea has been applied successfully for $\mathrm{SU}(2)$ using the Cayley transform \eqref{eq:cayley-transform}, resulting in a faster computational process for the Störmer--Verlet scheme \cite{wandelt2021geometric}. We will adapt this idea for $\mathrm{SU}(3)$ using the modified Cayley transform \eqref{modified Cayley}.

\subsection{Splitting methods based on the modified Cayley transform}
Replacing the matrix exponential in the exact flow \eqref{eq:link_update} by the modified Cayley transform \eqref{modified Cayley} results in a numerical approximation to the exact flow of convergence order $p=1$, as it can be regarded as the application of a Lie-Euler step \eqref{eq:Lie-Euler} with $\Psi = \caymod$. For this purpose, we need to compute $\mathrm{d}\caymod_{\Omega}^{-1}(A)$ at $\Omega = 0$.

\theorem{\label{theorem:auxODE}
The differential of $\caymod(\Omega)$ is given as
$$\left( \frac{\mathrm{d}}{\mathrm{d}\Omega} \caymod(\Omega) \right) A \bigg\vert_{\Omega = 0} = \Big( \mathrm{d}\caymod_{\Omega}(A) \Big) \caymod(\Omega).$$
The operator $\mathrm{d}\caymod_{\Omega}$ at $\Omega = 0$ is given as
\begin{align*}
    \mathrm{d} \caymod_{\Omega=0} (A) &= 2A, 
\intertext{and consequently its inverse reads}
    \mathrm{d} \caymod_{\Omega=0}^{-1}(A) &= \tfrac{1}{2} A.
\end{align*}
}
\proof{
Applying the Leibniz rule results in
\begin{align}\label{eq:diff_caymod}
    \left( \frac{\mathrm{d}}{\mathrm{d}\Omega} \caymod(\Omega)  \right) A &= \left( \frac{\mathrm{d}}{\mathrm{d}\Omega} (I - e^{-i\theta(\Omega)} \Omega)^{-1}\right) A (I + e^{i \theta(\Omega)} \Omega) + (I - e^{-i\theta(\Omega)}\Omega)^{-1} \left(\frac{\mathrm{d}}{\mathrm{d}\Omega} (I + e^{i\theta(\Omega)} \Omega) \right)A.
\end{align}
We are performing an auxiliary computation by considering the equation 
$$\left( \frac{\mathrm{d}}{\mathrm{d}\Omega} (I - e^{-i\theta(\Omega)} \Omega)^{-1} (I - e^{-i\theta(\Omega)}\Omega) \right) A = 0.$$
Applying the Leibniz rule to this equation, followed by a simple algebraic manipulation results in
\begin{equation}\label{eq:auxiliary_computation}
\left( \frac{\mathrm{d}}{\mathrm{d} \Omega}(I - e^{-i\theta(\Omega)}\Omega)^{-1} \right)A = -(I-e^{-i\theta(\Omega)}\Omega)^{-1} \left( \frac{\mathrm{d}}{\mathrm{d}\Omega} (I-e^{-i\theta(\Omega)} \Omega) \right)A (I-e^{-i\theta(\Omega)}\Omega)^{-1}.
\end{equation}
Inserting \eqref{eq:auxiliary_computation} into \eqref{eq:diff_caymod} yields
$$\left( \frac{\mathrm{d}}{\mathrm{d}\Omega} \caymod(\Omega) \right)A = (I - e^{-i\theta(\Omega)} \Omega)^{-1} \left[ -\left( \frac{\mathrm{d}}{\mathrm{d} \Omega} (I - e^{-i\theta(\Omega)} \Omega) \right) A + \left(\frac{\mathrm{d}}{\mathrm{d} \Omega} (I + e^{i\theta(\Omega)}\Omega)\right) A \caymod(\Omega)^{-1} \right] \caymod(\Omega),$$
so that 
\begin{align*}
    \mathrm{d} \caymod_{\Omega}(A) = (I - e^{-i\theta(\Omega)}\Omega)^{-1} \left[ \left(\frac{\mathrm{d}}{\mathrm{d}\Omega} (I - e^{-i\theta(\Omega)}\Omega)\right) A + \left( \frac{\mathrm{d}}{\mathrm{d}\Omega} (I + e^{i\theta(\Omega)} \Omega) \right)A \caymod(\Omega)^{-1} \right].
\end{align*}
By applying the chain rule and inserting $\Omega = 0$, we immediately get
\begin{align*}
    -\left(\frac{\mathrm{d}}{\mathrm{d}\Omega} (I - e^{-i\theta(\Omega)}\Omega)\right) A \bigg\vert_{\Omega = 0} = \left( \frac{\mathrm{d}}{\mathrm{d}\Omega} (I + e^{i\theta(\Omega)} \Omega) \right)A\bigg\vert_{\Omega = 0} = A,
\end{align*}
and with $\caymod(0) = I$ finally 
$$ \mathrm{d} \caymod_{\Omega=0} (A) = 2A.$$
Hence, its inverse is given by 
$$\mathrm{d} \caymod_{\Omega=0}^{-1} (A) = \tfrac{1}{2}A,$$
proving the statements of the theorem.
\qedhere}

\remark{It holds $\mathrm{d}\cay_{\Omega = 0}^{-1}(A) = \mathrm{d}\caymod_{\Omega = 0}^{-1}(A) = A/2$.}\normalfont\medskip

Based on Theorem \ref{theorem:auxODE}, the Lie-Euler step \eqref{eq:Lie-Euler} with $\Psi = \caymod$ becomes 
\begin{align}\label{eq:Lie-Euler_caymod}
    Y_1 = \caymod\left( \tfrac{h}{2} A(t_0) \right) Y_0.
\end{align}
To clarify the use of the numerical approximation \eqref{eq:Lie-Euler_caymod} instead of the exact flow \eqref{eq:link_update}, we will denote a link update with the modified Cayley transform with $\phi_{a_i h}^{\{1\}}$ in the definition of the splitting methods instead of $\varphi_{a_i h}^{\{1\}}$.

\theorem{Replacing the exact flow \eqref{eq:link_update} by the Lie-Euler step \eqref{eq:Lie-Euler_caymod} affects neither the time-reversibility nor the volume-preservation of the splitting method.}
\proof{Since the momentum updates \eqref{eq:momentum_update} remain unchanged, we only have to investigate the approximated link updates \eqref{eq:link_update} using \eqref{eq:Lie-Euler_caymod}.\medskip

\noindent\emph{Time-reversibility.} To prove the time-reversibility of the Lie-Euler step \eqref{eq:Lie-Euler_caymod}, denoted by $\phi_h^{\{1\}}$, we have to show that
$$\rho \circ \phi_h^{\{1\}} \circ \rho \circ \phi_h^{\{1\}}(A_0,Y_0) = (A_0,Y_0),$$
with $\rho(A_0,Y_0) = (-A_0,Y_0)$ changing the sign of momenta. A straight-forward application of $\phi_h^{\{1\}}$, followed by applying $\rho$ yields $\left(-A_0,\caymod(\tfrac{h}{2}A_0)Y_0\right)$. Again applying $\phi_h^{\{1\}}$ and using the identity $\caymod(A) \caymod(-A) = I$ results in $(-A_0,Y_0)$. A final application of $\rho$ yields the initial values $(A_0,Y_0)$ again, showing the time-reversibility of the link update.\medskip

\noindent\emph{Volume-preservation.} For volume-preservation of the link update $\phi_h^{\{1\}}$, we have to show that
$$\left\lvert \det \frac{\partial \phi_h^{\{1\}}(A_0,Y_0)}{\partial(A_0,Y_0)} \right\rvert = 1.$$
The Jacobian reads 
\begin{align*}
    \frac{\partial \phi_h^{\{1\}}(A_0,Y_0)}{\partial(A_0,Y_0)} = \begin{pmatrix}
        I & 0 \\
        \mathrm{d}\caymod_{A_0}(Y_0) \caymod(A_0) & \caymod(A_0)
    \end{pmatrix}.
\end{align*}
Since $\caymod : \mathfrak{su}(3) \to \mathrm{SU}(3)$, it holds $\det(\caymod(A_0)) = 1$ and thus also the overall Jacobian has determinant 1, proving the volume-preservation. \qedhere
}

\subsection{Examples}
In this section, we will present common splitting methods used in lattice QCD simulations and discuss which splitting methods still work with the desired order of convergence when replacing the matrix exponential by the modified Cayley transform \eqref{modified Cayley}.

\subsubsection{Splitting methods of convergence order two}
Since the Lie-Euler step \eqref{eq:Lie-Euler} with $\Psi = \caymod$ is a convergent numerical scheme, all symmetric splitting methods are at least of convergence order two. Consequently, the convergence order of all splitting methods of order $p=2$ will not be affected by using the modified Cayley transform \eqref{modified Cayley} instead of the matrix exponential. 

\example[Störmer--Verlet method \cite{hairer2003geometric}]{
The Cayley-version of the Störmer--Verlet algorithm in velocity version reads 
\begin{align*}
    \Phi_h = \varphi_{h/2}^{\{2\}} \circ \phi_h^{\{1\}} \circ \varphi_{h/2}^{\{2\}}.
\end{align*}
Analogously, the position version reads
\begin{align*}
    \Phi_h = \phi_{h/2}^{\{1\}} \circ \varphi_h^{\{2\}} \circ \phi_{h/2}^{\{1\}}.
\end{align*}}\normalfont

By using more stages per time step, more efficient integrators can be derived. 
\example[Second-order minimum norm (2MN) scheme \cite{omelyan2003symplectic}]{Let
$$\lambda = \frac{1}{2} - \frac{(2 \sqrt{326} + 36)^{1/3}}{12} + \frac{1}{6(2\sqrt{326} + 36)^{1/3}} \approx 0.1932.$$
Then, the velocity version of the five-stage decomposition reads
\begin{align}\label{eq:OMF2_velocity}
    \Phi_h = \varphi_{\lambda h}^{\{2\}} \circ \phi_{h/2}^{\{1\}} \circ \varphi_{(1-2\lambda)h}^{\{2\}} \circ \phi_{h/2}^{\{1\}} \circ \varphi_{\lambda h}^{\{2\}}.
\end{align}
Similarly, one obtains for the position-like algorithm
\begin{align}\label{eq:OMF2_position}
    \Phi_h = \phi_{\lambda h}^{\{1\}} \circ \varphi_{h/2}^{\{2\}} \circ \phi_{(1-2\lambda)h}^{\{1\}} \circ \varphi_{h/2}^{\{2\}} \circ \phi_{\lambda h}^{\{1\}}.
\end{align}
}\normalfont
\remark[Optimality of 2MN]{The free parameter $\lambda$ has been derived in order to minimize the norm of the principal error term for $\Psi = \exp$. Since the principal error term changes due to the usage of $\caymod$, this value is not optimal anymore.}\normalfont

\subsubsection{Splitting methods of higher order}\label{sec:Higher-order}
There exist many different approaches to derive splitting methods of order $p>2$, for example, direct decomposition algorithms \cite{omelyan2003symplectic}, force-gradient integrators \cite{omelyan2003symplectic,kennedy2009force} as well as their Hessian-free versions \cite{yin2011improving,schaefers2024hessianfree,shcherbakov2017adapted}, and composition schemes \cite{omelyan2002construction,kahan1997composition,suzuki1990fractal,yoshida1990construction}.
The derivation of direct decomposition algorithms and force-gradient integrators are assuming that the flows of the subsystems are computed exactly. In general, since the Lie-Euler step \eqref{eq:Lie-Euler_caymod} is convergent of order $p=1$, an order reduction will appear if the splitting method is of order $p>2$ ($p=2$ still works due to symmetry).
Composition schemes, however, still allow for deriving splitting methods of arbitrarily high convergence order while using the modified Cayley transform \eqref{modified Cayley} instead of the matrix exponential. Consider a splitting method $\Phi_h$ of order $p$, for example, one of those from the previous section. Then, the composition method
\begin{align*}
    \Gamma_h = \Phi_{\gamma_r h} \circ \ldots \circ \Phi_{\gamma_1 h}
\end{align*}
has convergence order $p + 2$, if the following conditions are satisfied.
\begin{subequations}\label{eq:composition_system}
\begin{align}
    \sum_j^r \gamma_j &= 1, & &\text{(full time step)}, \\
    \sum_j^r \gamma_j^{p+1} &= 0, & &\text{(vanishing leading error term)}, \\
    \gamma_{r+1-j} &= \gamma_j, \quad j=1,\ldots,r, & &\text{(symmetry)}.
\end{align}
\end{subequations}
\example[Yoshida's triple-jump \cite{yoshida1990construction}]{The smallest value of $r$ that allows for a real solution of \eqref{eq:composition_system} is $r=3$. The unique solution is given by
\begin{align}\label{eq:Yoshida_weights}
    \gamma_1 = \gamma_3 = 1/(2 - 2^{1/(p+1)}), \quad \gamma_2 = 1 - 2\gamma_1.
\end{align}}\normalfont

The investigations done in \cite{omelyan2003symplectic} show that the weights \eqref{eq:Yoshida_weights} do not result in efficient composition schemes. By increasing $r$, one is able to obtain more efficient composition schemes, for example via the following composition technique. 
\example[Suzuki's fractals]{For $r=5$, the best solution of \eqref{eq:composition_system} is given by 
\begin{align}\label{eq:Suzuki_weights}
    \gamma_1 = \gamma_2 = \gamma_4 = \gamma_5 = 1/(4 - 4^{1/(p+1)}), \quad \gamma_3 = 1-4\gamma_1. 
\end{align}}\normalfont

Starting from a splitting method based on the modified Cayley transform \eqref{modified Cayley}, e.g.\ the five-stage algorithm \eqref{eq:OMF2_velocity}, Yoshida's triple-jump \eqref{eq:Yoshida_weights} and Suzuki's fractals \eqref{eq:Suzuki_weights} are the most promising techniques to obtain splitting methods of order $p=4$. One could apply these composition techniques iteratively up to the desired convergence order. However, the number of applications of the underlying splitting method grows drastically. If one aims for splitting methods of order $p \geq 6$, advanced composition techniques \cite{omelyan2002construction} allow for deriving composition schemes of higher order with less stages. The special case of starting with a base scheme of order $p=2$ has been investigated in \cite{kahan1997composition}.

\example[6th order advanced composition (AC6) \cite{kahan1997composition}]{Starting from a base method of order two, $r=7$ applications of the base scheme with weights
\begin{align*}
    \gamma_1 &= \gamma_7 = 0.78451361047755726382, &
    \gamma_2 &= \gamma_6 = 0.23557321335935813368, \\
    \gamma_3 &= \gamma_5 = -1.1776799841788710069, &
    \gamma_4 &=  1.3151863206839112189,
\end{align*}
are necessary to obtain a composition scheme $\Gamma_h$ of convergence order six.}\normalfont 

For more advanced composition schemes up to order $p=10$, we refer to \cite{kahan1997composition}.

\remark{
In lattice QCD, one frequently works with the traceless and Hermitian momenta $P_{x,\mu}$ rather than the Lie algebra elements $iP_{x,\mu}$. Let $P_{x,\mu} \in \mathbb{R}^{3 \times 3}$ be traceless and Hermitian. With
\begin{subequations}\label{eq:modified_cayley_P}
\begin{align*}
    \theta(P_{x,\mu}) = \begin{cases}
        \pi/2, & \gamma(P_{x,\mu})=0, \\
        \cos^{-1}\left(\tfrac{1}{2}\left(\gamma(P_{x,\mu})^{-1} - \sqrt{\gamma(P_{x,\mu})^{-2} + 1}\right)\right), & \gamma(P_{x,\mu}) > 0, \\
        \cos^{-1}\left(\tfrac{1}{2}\left(\gamma(P_{x,\mu})^{-1} + \sqrt{\gamma(P_{x,\mu})^{-2} + 1}\right)\right), & \gamma(P_{x,\mu}) < 0,
    \end{cases}
\end{align*} 
and $\gamma(P_{x,\mu}) = 4 \det(P_{x,\mu})/\mathrm{tr}(P_{x,\mu}^2)$ so that $\theta(P_{x,\mu}) \in (\pi/3,2\pi/3)$, the mapping 
\begin{align*}
    \left( I + e^{-i\theta(P_{x,\mu})}P_{x,\mu} \right)^{-1} \left(I + e^{i\theta(P_{x,\mu})}P_{x,\mu}\right)
\end{align*}
maps $P_{x,\mu}$ into the special unitary group $\mathrm{SU}(3)$.
\end{subequations}
}\normalfont

\section{Numerical results}\label{sec:numerical_results} 
\begin{figure}[htbp!]
    \centering
    \includegraphics[]{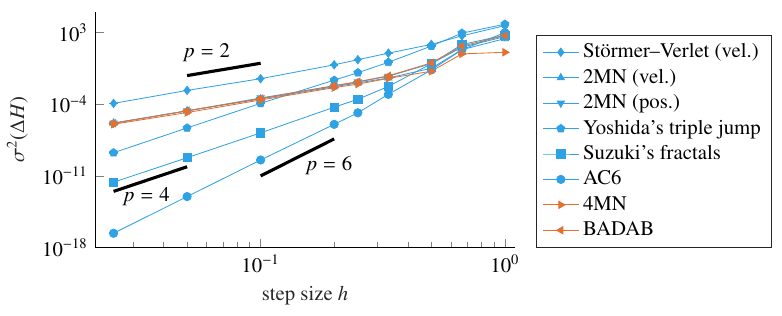}
    \caption{Variance of $\Delta H$ vs.\ step size $h$ for different decomposition algorithms using the modified Cayley transform $\Psi = \caymod$. For an integrator of order $p$, the variance $\sigma^2(\Delta H)$ scales with order $2p$. Whereas the order of second-order integrators and composition schemes (blue lines) remains unchanged, higher-order direct decomposition algorithms and force-gradient integrators are affected by an order reduction (orange lines). For all composition techniques, the velocity version of the Störmer--Verlet method has been used as the underlying base scheme.}
    \label{fig:caymod_order}
\end{figure}

For the numerical tests, we consider a similar setting as in \cite{wandelt2021geometric} where the Cayley transform has been investigated for simulations in an $\mathrm{SU}(2)$ Yang-Mills theory using the HMC algorithm. 
We perform numerical simulations in an $\mathrm{SU}(3)$ Yang-Mills theory using the HMC algorithm on a two-dimensional lattice of size $V= 32 \times 32$ with lattice spacing $a = 1$. Here, the Hamiltonian takes the form \eqref{eq:Hamiltonian} with $S([U])$ denoting the Wilson gauge action. For the gauge coupling, we choose $\beta = 2.0$.
In a first simulation, we computed 1000 trajectories of length $\tau = 2.0$ for varying step sizes $h$ and all integrators introduced in Section \ref{sec:splitting_methods} with $\Psi = \caymod$. The results are depicted in Figure \ref{fig:caymod_order}, confirming the theoretical findings on the convergence order of the Cayley-based integrators. 

\begin{figure}[tbh]
    \centering
    \includegraphics[]{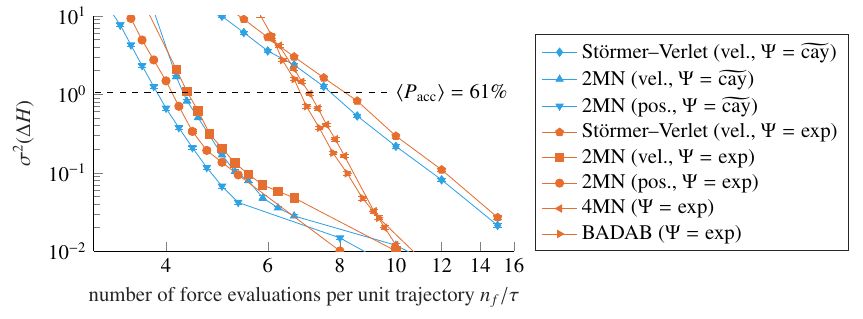}
    \caption{Variance of $\Delta H$ vs.\ number of force evaluations per unit trajectory trajectory $n_f / \tau$ for second-order Cayley-based integrators (blue lines) and a selection of commonly used decomposition algorithms using the matrix exponential (orange lines).}
    \label{fig:caymod_efficiency}
\end{figure}

\begin{figure}[tbh]
    \centering
    \includegraphics[]{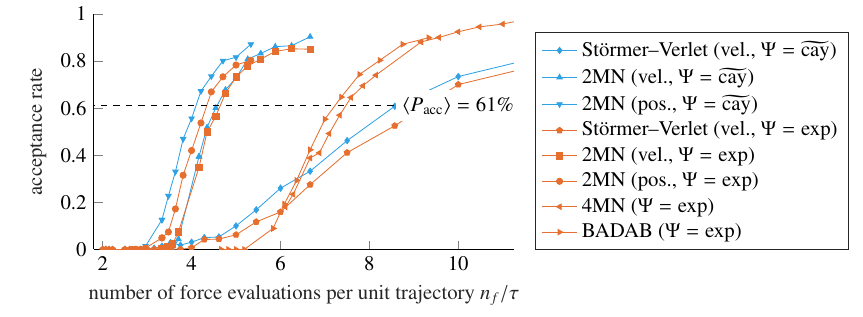}
    \caption{Acceptance rate vs.\ number of force evaluations per unit trajectory $n_f / \tau$ for second-order Cayley-based integrators (blue lines) and a selection of commonly used decomposition algorithms using the matrix exponential (orange lines).}
    \label{fig:caymod_acceptance}
\end{figure}

On this small lattice, second-order algorithms turn out to be the most efficient ones \cite{takaishi2006}. Hence we focus on a comparison of Cayley-based integrators of second order and commonly used decomposition algorithms based on the matrix exponential. In HMC, one is interested in good energy conservation, ensuring a high acceptance rate in the Metropolis step, while minimizing the computational cost. 
Here, the choice of the integrator in the molecular dynamics step plays a crucial role. The optimal acceptance rate for an integrator of order $p=2$ is $\langle P_{\mathrm{acc}}\rangle_\mathrm{opt} = \exp(-1/2) \approx 61\%$ \cite{TAKAISHI20006}. 
For log-normal distributed $\Delta H$, the expected acceptance rate can be determined based on the (empirical) variance $\sigma^2(\Delta H)$ and is given by $\langle P_{\mathrm{acc}}\rangle = \mathrm{erfc}(\sqrt{\sigma^2(\Delta H)/8})$ with $\mathrm{erfc}$ the complementary error function. 
In our Matlab implementation, the modified Cayley transform is slightly less expensive to evaluate than the exponential map (using the analytical formula derived in \cite{Kaiser_ExpForSU3}).
Depending on the total number of links, evaluating the modified Cayley transform is approximately 10-20\% faster.\footnote{In \cite{wandelt2021geometric}, the Störmer--Verlet method using the Cayley transform was $\approx 4.5$ times faster than the version based on the matrix exponential. In~\cite{wandelt2021geometric}, however, the matrix exponential has been computed via the significantly slower matrix decomposition method described in \cite{Moler_VanLoan}.}
In a preliminary, non-optimized implementation of the modified Cayley transform in openQCD \cite{openQCD}, there were no significant differences in the computational cost compared to the implementation of the matrix exponential.
Given that the overall computational cost of the HMC algorithm is governed by the number of force evaluations $n_f$, these potentially minor variations in the evaluation costs of the local parameterizations become negligible. 
The critical question is whether the application of the modified Cayley transform leads to an increased acceptance probability, enabling the utilization of larger step sizes and consequently reducing the number of force evaluations per trajectory of length $\tau$.
We thus aim at minimizing the total number of force evaluations per unit trajectory $n_f/\tau$ while achieving the optimal acceptance rate of $61\%$. For this purpose, we start from a thermalized configuration and compute 1000 trajectories of varying trajectory length $\tau$ and step size $h=\tau/2$.  
In Figure \ref{fig:caymod_efficiency}, the variance $\sigma^2(\Delta H)$ is depicted for different values of $n_f/\tau$. The horizontal line indicates the optimal acceptance rate of $61\%$.
Similarly, Figure \ref{fig:caymod_acceptance} shows the acceptance rate vs.\ $n_f/\tau$.
Both figures indicate that the five-stage decomposition algorithm \eqref{eq:OMF2_position} is the most efficient splitting method among all integrators under investigation. This coincides with the investigations made in \cite{takaishi2006}. Furthermore, the results emphasize that the use of second-order decomposition algorithms with $\Psi = \caymod$ results in a more efficient computational process. Especially for the most efficient integrator \eqref{eq:OMF2_position} we are able to see significant differences in the acceptance rate. 

%
For smaller step sizes, the exponential-based integrators become better than their Cayley-based counterparts.
This is not surprising since the use of the modified Cayley transform introduces an additional error source by approximating the exact flow \eqref{eq:link_update}.  
Despite the additional error source, the Cayley-versions perform better for acceptance rates close to $\langle P_{\mathrm{acc}}\rangle_\mathrm{opt} = 61\%$, i.e.\ in the domain of interest. In this region, the decomposition algorithms are close to the border of their stability domain. Using Neumann series, one can expand the modified Cayley transform in \eqref{eq:Lie-Euler_caymod} and gets 
\begin{align*}
    \caymod(\tfrac{h}{2} A) &= (I - \tfrac{h}{2} e^{-i\theta} A)^{-1} (I + \tfrac{h}{2} e^{i\theta} A) = I + \tfrac{h}{2} 2 \cos(\theta) A + \tfrac{h^2}{4} (1+ e^{-2i\theta}) A^2 + \mathcal{O}(h^3).
\end{align*}
Since $\lvert \cos(\theta)\rvert \le 1$ and $\lvert e^{-2i\theta}\rvert = 1$, the modified Cayley transform shows a damping behavior in the leading terms compared to the matrix exponential 
$$\exp(hA) = I + hA + \frac{h^2}{2}A^2 + \mathcal{O}(h^3).$$
The numerical results give evidence that the damping of the modified Cayley transform has a stabilizing effect that results in higher acceptance rates in the HMC algorithm.
All in all, the numerical results emphasize that the use of the modified Cayley transform as the local parameterization is beneficial for second-order integrators.

\remark[Projection-based modification of the Cayley transform]{
An alternative approach to make the Cayley transform suitable for $\mathrm{SU}(3)$ is given by the projection
\begin{equation}\label{eq:cay_projection}
\widehat{\cay}(\Omega) := \det(\cay(\Omega))^{-1/3} \cay(\Omega). 
\end{equation}
The mappings \eqref{eq:cay_projection} and \eqref{modified Cayley} have similar evaluation costs. However, an expansion of the projection \eqref{eq:cay_projection} reveals 
\begin{align*}
    \widehat{\cay}(\tfrac{h}{2} A) = \det(\cay(\Omega))^{-1/3} \cdot \left[ I + hA + \frac{h^2}{2} A^2 + \frac{h^3}{4} A^3 + \mathcal{O}(h^4) \right].
\end{align*} 
With $\lvert\det(\cay(\Omega))\rvert = 1$, the projection does not have the beneficial damping behavior like the modified Cayley transform \eqref{modified Cayley}. 
Instead, the expansion even has a larger constant in the $\mathcal{O}(h^3)$ term compared to the matrix exponential. 
Consequently, the use of the projection \eqref{eq:cay_projection} will not result in beneficial stability properties. 
}\normalfont

\section{Conclusion and outlook}
In this paper, we proposed a modification to the Cayley transform that defines a local parameterization for the special unitary group $\mathrm{SU}(3)$. We discussed how to use the modified Cayley transform instead of the matrix exponential inside splitting methods and proved that the replacement does affect neither the time-reversibility nor the volume-preservation of the splitting method. For composition schemes, the convergence order is not affected by the modified Cayley transform, allowing for the derivation of splitting methods of arbitrarily high convergence order that are based on the modified Cayley transform. More advanced decomposition algorithms and force-gradient integrators of order $p>2$ suffer from an order reduction. Numerical results highlight that the modified Cayley transform allows for the derivation of more efficient decomposition algorithms of convergence order two.

Direct decomposition methods and force-gradient integrators \cite{omelyan2003symplectic} keep their convergence order if the exact flow of the link update \eqref{eq:link_update} is approximated using a Lie group method of the same convergence order, e.g.\ based on the Munthe-Kaas approach \cite{munthe1998runge}. 
In general, the Munthe-Kaas approach demands the auxiliary ODE
$$ \dot{\Omega} = \mathrm{d} \caymod_\Omega^{-1}(A),$$
also in the case $\Omega \neq 0$. 
Since the angle $\theta(\Omega)$ depends on $\Omega$, the derivative is more complicated to derive than for the traditional Cayley transform \eqref{eq:cayley-transform}. 
In addition, the Munthe-Kaas approach may prove advantageous in more complicated scenarios where the Hamiltonian is not separable, such as when considering the Riemannian manifold HMC (RMHMC) algorithm \cite{Nguyen:20226k,Jung:2024o1,Fields:2025El,Kennedy:2025X2}, which presents a promising approach for overcoming critical slowing down in modern lattice computations.
A preliminary step in designing numerical integration schemes for the RMHMC algorithm involves deriving symplectic and time-reversible Lie group integrations based on the exponential map, for instance, by extending the symplectic partitioned Lie group methods proposed in \cite{bogfjellmo2016high}.
Subsequently, a modification of this framework by employing the modified Cayley transform as the local parameterization may facilitate the derivation of more efficient numerical integration schemes.
Furthermore, we will test the proposed integrators based on the modified Cayley transform in large-scale lattice QCD simulations.

\section*{Declaration of competing interest}
The authors declare that they have no known competing financial interests or personal relationships that could have appeared to influence the work reported in this paper.

\section*{Data availability}
The numerical tests have been performed using a MATLAB implementation that is opensource and available online at \url{https://github.com/KevinSchaefers/pure-gauge_SU3}.

\section*{Acknowledgements}
This work is supported by the German Research Foundation (DFG) research unit FOR5269 "Future methods for studying confined gluons in QCD" and by the STRONG-2020 project, funded by the European Community Horizon 2020 research and
innovation programme under grant agreement 824093.
\appendix

\section{Gell-Mann matrices}\label{app:Gell-Mann}
A basis of traceless and Hermitian matrices $A \in \mathbb{C}^{3 \times 3}$ is given by the Gell--Mann matrices, which are given by
\begin{align}\label{eq:Gell-Mann}
    \lambda_1 &= \begin{pmatrix} 0 & 1 & 0 \\ 1 & 0 & 0 \\ 0 & 0 & 0\end{pmatrix}, & 
    \lambda_2 &= \begin{pmatrix} 0 & -i & 0 \\ i & 0 & 0 \\ 0 & 0 & 0\end{pmatrix}, & 
    \lambda_3 &= \begin{pmatrix} 1 & 0 & 0 \\ 0 & -1 & 0 \\ 0 & 0 & 0\end{pmatrix}, &
    \lambda_4 &= \begin{pmatrix} 0 & 0 & 1 \\ 0 & 0 & 0 \\ 1 & 0 & 0\end{pmatrix}, \notag \\ 
    \lambda_5 &= \begin{pmatrix} 0 & 0 & -i \\ 0 & 0 & 0 \\ i & 0 & 0\end{pmatrix}, & 
    \lambda_6 &= \begin{pmatrix} 0 & 0 & 0 \\ 0 & 0 & 1 \\ 0 & 1 & 0\end{pmatrix}, &
    \lambda_7 &= \begin{pmatrix} 0 & 0 & 0 \\ 0 & 0 & -i \\ 0 & i & 0\end{pmatrix}, & 
    \lambda_8 &= \frac{1}{\sqrt{3}}\begin{pmatrix} 1 & 0 & 0 \\ 0 & 1 & 0 \\ 0 & 0 & -2\end{pmatrix}.
\end{align}
Consequently, the matrices $i\lambda_j,\, j=1,\ldots,8$, build a basis of the Lie algebra $\mathfrak{su}(3)$.
\section{Fourth-order decomposition algorithms}
Composition techniques should be used for the construction of algorithms of extremely high orders only, where the derivation via direct decomposition results in an unresolvable numerical problem \cite{omelyan2003symplectic}. We therefore highlight two fourth-order decomposition algorithms of order $p=4$ with coefficients derived in \cite{omelyan2003symplectic} and that are frequently used in lattice QCD simulations.\medskip 

\noindent\emph{Fourth-order minimum norm (4MN) scheme.}\ 
By minimizing the norm of the leading error coefficients, the most efficient non-gradient algorithm of order $p=4$ is given by the eleven-stage decomposition
\begin{align*}
    \Phi_h = \varphi^{\{2\}}_{b_1 h} \circ \varphi^{\{1\}}_{a_2 h} \circ \varphi^{\{2\}}_{b_2 h} \circ \varphi^{\{1\}}_{a_3 h} \circ \varphi^{\{2\}}_{(1/2 - b_1 - b_2) h} \circ \varphi^{\{1\}}_{(1 - 2(a_2+a_3)) h} \circ \varphi^{\{2\}}_{(1/2 - b_1 - b_2) h} \circ \varphi^{\{1\}}_{a_3 h} \circ \varphi^{\{2\}}_{b_2 h} \circ \varphi^{\{1\}}_{a_2 h} \circ \varphi^{\{2\}}_{b_1 h}
\end{align*}
with parameters 
\begin{align*}
    a_2 &= 0.253978510841060, & a_3 &= -0.032302867652700, &
    b_1 &= 0.083983152628767, & b_2 &= 0.682236533571909.
\end{align*}\medskip

\noindent\emph{Hessian-free force-gradient integrator (BADAB)} \cite{yin2011improving,schaefers2024hessianfree}.\  
The Hessian-free variant of the most efficient force-gradient integrator with five stages reads 
\begin{align*}
    \Phi_h = \varphi_{h/6}^{\{2\}} \circ \varphi^{\{1\}}_{h/2} \circ \tilde{\varphi}^{\{2\}}_{2h/3}  \circ \varphi^{\{1\}}_{h/2} \circ \varphi_{h/6}^{\{2\}},
\end{align*}
where $\tilde{\varphi}$ denotes an approximated force-gradient step including a temporary link update. For details, we refer to \cite{schaefers2024hessianfree,Schäfers:2025BT,schaefers2025stability}.

For both methods, replacing the matrix exponential by the modified Cayley transform, i.e.\ approximating the link updates by a Lie-Euler step \eqref{eq:Lie-Euler_caymod} with $\Psi = \caymod$, results in algorithms that are convergent of order two only. 

\bibliographystyle{elsarticle-num} 
\bibliography{mybibfile}
\end{document}